\documentclass[5p,twocolumn,times,number]{elsarticle}

\usepackage{graphicx}
\usepackage{amsmath}   

\begin{document}

\begin{frontmatter}

\title{Upgrades of the ATLAS Muon Spectrometer with sMDT Chambers}

\author[add1]{C.~Ferretti}
\author[add2]{H.~Kroha\corref{cor}}
\ead{kroha@mppmu.mpg.de}
\author{on behalf of the ATLAS Muon Collaboration}

\cortext[cor]{Corresponding author}

\address[add1]{University of Michigan, Ann Arbor, USA}
\address[add2]{Max-Planck-Institut f\"ur Physik, Munich, Germany}

\begin{abstract}
With half the drift-tube diameter of the Monitored Drift Tube (MDT) chambers of the ATLAS muon spectrometer and 
otherwise unchanged operating parameters, small-diameter Muon Drift Tube (sMDT) chambers provide an order of magnitude 
higher rate capability and can be installed in detector regions where MDT chambers do not fit. 
The chamber assembly time has been reduced by a factor of seven to one working day and the sense wire 
positioning accuracy improved by a factor of two to better than ten microns. Two sMDT chambers have been installed 
in ATLAS in 2014 to improve the momentum resolution in the barrel part of the spectrometer. 
The construction of an additional twelve chambers covering the feet regions of the ATLAS detector has started. 
It will be followed by the replacement of the MDT chambers at the ends of the barrel inner layer by sMDTs improving 
the performance at the high expected background rates and providing space for additional RPC trigger chambers. 
\end{abstract}


\end{frontmatter}

\section{Introduction}

The goals of the ATLAS muon detector upgrades are to increase the acceptance for precision muon momentum measurement and triggering 
and to improve the rate capability of the muon chambers in the high-background regions for increasing instantaneous luminosity of the LHC.
With 15~mm drift-tube diameter instead of the 30~mm of the Monitored Drift Tube (MDT) chambers~\cite{MDT} of the ATLAS muon spectrometer and otherwise
unchanged operating parameters, the small-diameter Muon Drift Tube (sMDT) chambers~\cite{sMDT} provide an order of magnitude 
higher rate capability and can be installed in detector regions where MDT chambers do not fit. Since they use the same Ar:CO$_2$ (93:7)
gas mixture at 3 bar and the same readout electronics as the MDTs, the sMDT chambers can be easily integrated into the infrastructure, 
optical alignment system, data acquisition and track reconstruction software of the ATLAS muon spectrometer. Platforms for the alignment sensors
are mounted with high accuracy with respect to the sense wires during chamber assembly allowing for absolute chamber misalignment corrections.  

Two sMDT chambers (called BME) covering acceptance gaps in the bottom sector of the barrel part of the muon spectrometer 
have been constructed and installed in spring 2014. The construction of 12 additional sMDT chambers (called BMG) to be installed in the
feet of the ATLAS detector in the 2016/17 winter shutdown of the LHC has started at the end of last year. The design of sMDT chambers integrated with new RPC 
trigger chambers is in progress for the replacement of 16 MDT chambers at the ends of the inner barrel layer in the next long LHC shutdown  
with the goal of improving the muon tracking and trigger performance at high background rates.  
All the sMDT chambers mentioned consist of two quadruple layers of drift tubes separated by a spacer and support frame.
Already at the LHC design luminosity the ATLAS muon detectors are exposed to unprecedentedly high background rates of $\gamma$-rays and neutrons 
from energetic particle interactions in the detector and beam shielding, of up to 500~Hz/cm$^2$. For future upgrades of the LHC, 
the background rates are expected to increase by up to an order of magnitude roughly proportional to the instantaneous luminosity. 

\section{sMDT Chamber Technology and Performance}

In contrast to the MDT chambers, the sMDT chambers are constructed from standard industrial aluminum tubes with 0.4 mm wall thickness.
The drift tubes are assembled and tested for gas tightness, high-voltage stability and wire tension using semi-automated facilities 
in a temperature controlled clean room at a typical rate of 100 tubes per day.
A new chamber assembly method (see \cite{sMDT}) has been developed which allows for the gluing of a chamber within one working day.
The endplug design of the sMDT chambers~\cite{sMDT} is not only cheaper and more reliable than
for the MDT chambers but also provides higher wire positioning accuracy. In addition, the wire positions
in the endplugs can be measured directly with a few micron precision with a coordinate measuring machine 
via the external reference surfaces on the endplugs which are used for the positioning of the tubes in the chamber.
At the same time, the alignment platform positions are measured with respect to the sense wire positions. 
The signal wires are positioned in the endplugs with respect to these reference surfaces, which are    
on the same machined brass inserts of the endplugs which locate the wires, with better than 5 micron precision. 
The plastic insulator material of the injection moulded endplugs is not susceptible to cracking and shows no outgassing. 
The modular on-chamber gas distribution system uses injection moulded gas connectors 
consisting of the same plastic material as the endplugs.  

Due to the smaller tube diameter and the 3.8 times shorter maximum drift time, the rate capability of the sMDT chambers is increased by about an order 
of magnitude compared to the MDT chambers using the same readout electronics. Space charge effects degrading the spatial resolution are
almost eliminated up to the maximum expected background rates. The resolution and muon detection efficiency is rather limited by signal pile-up
effects of the current readout electronics and can be further improved~\cite{elba3}.

\vspace{-6mm}
\section{BME Chambers}

\begin{figure}
\vspace{-40mm}\hspace{-10mm}
\includegraphics[width=1.2\linewidth]{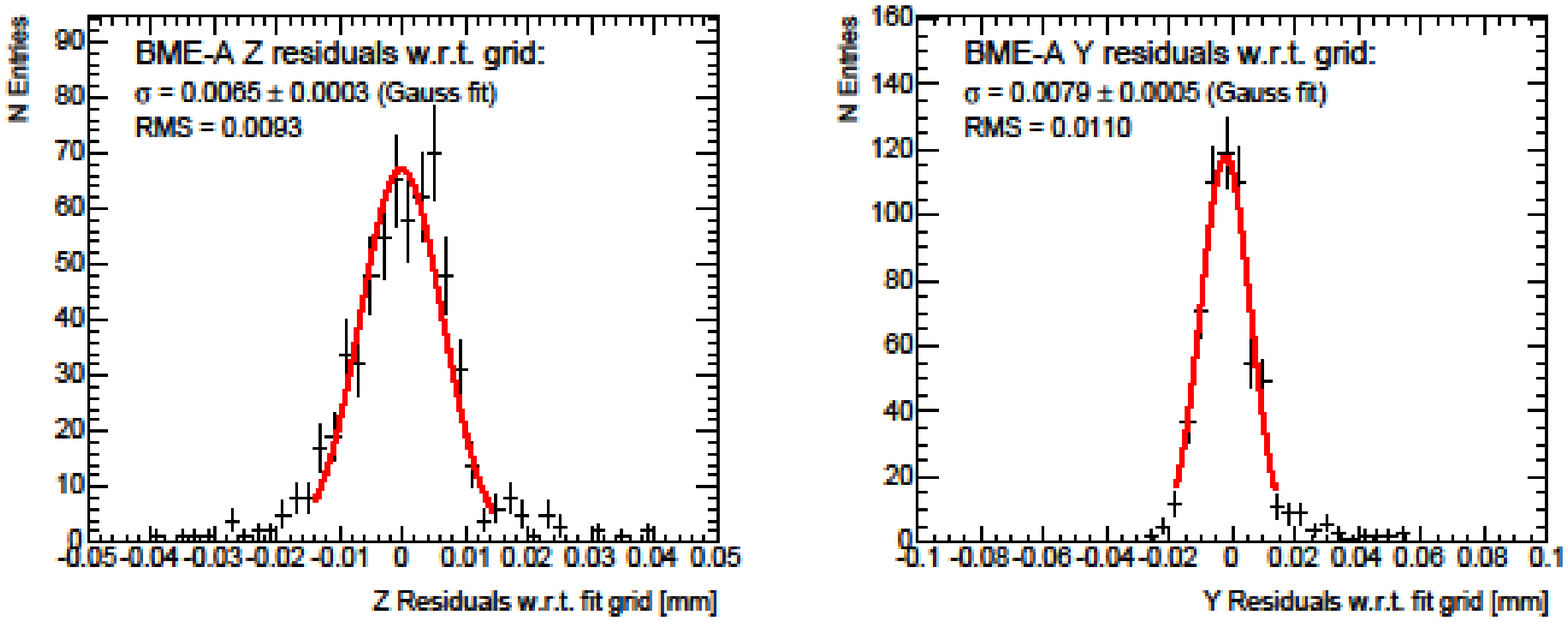}

\vspace{-27mm}
\caption{Residual distributions of the BME wire positions at the two tube ends from CMM measurements (y perpendicular
to the chamber plane).} 
\label{fig:BMEA_wirepos}
\end{figure}

\begin{figure}
\centering
\includegraphics[width=0.9\linewidth]{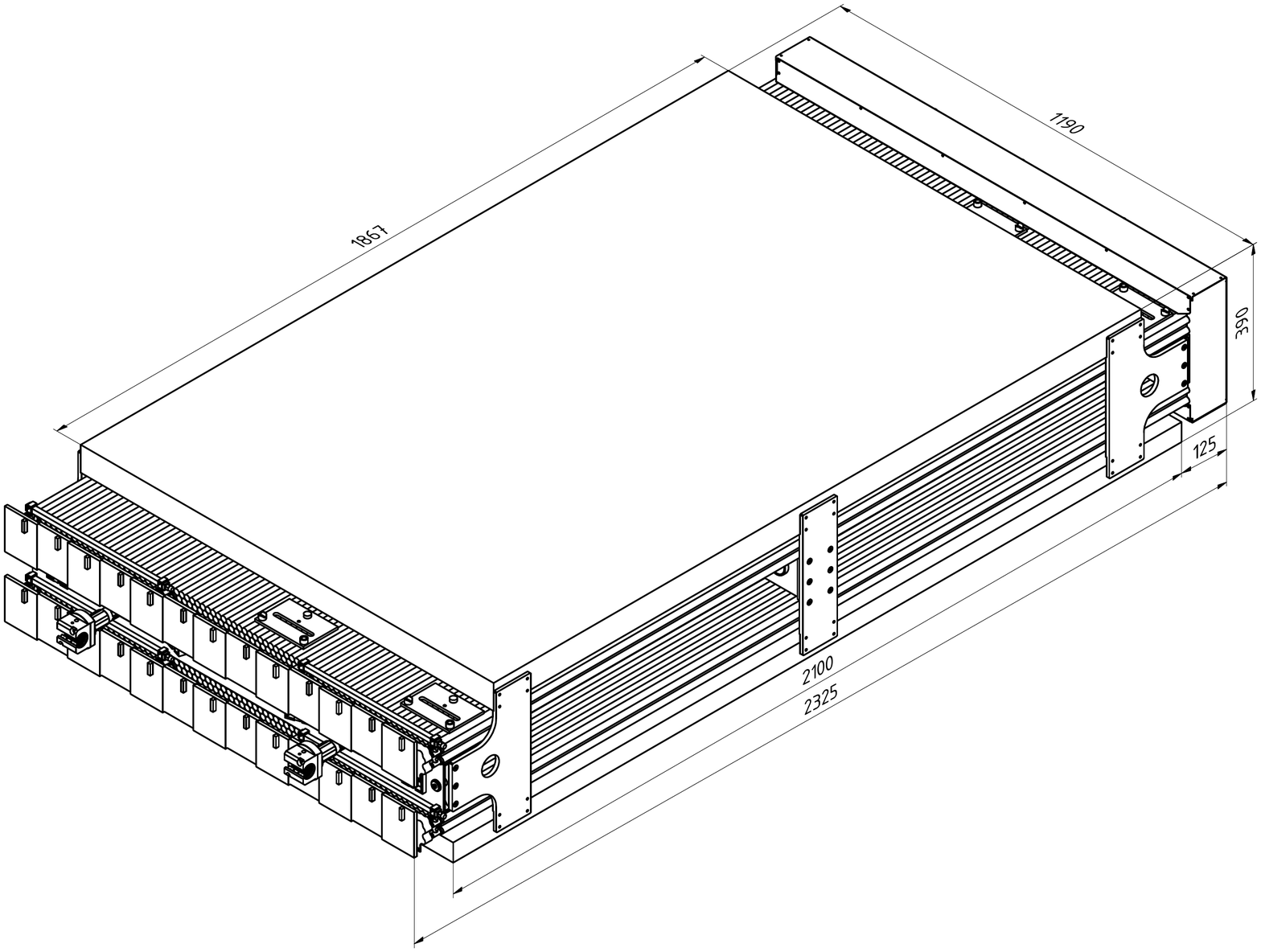}
\caption{Layout of a BME chamber with an RPC chamber mounted on top and bottom.}
\label{fig:BMEdesign}
\end{figure}

\begin{figure}
\vspace{-21mm}
\centering
\includegraphics[width=0.65\linewidth]{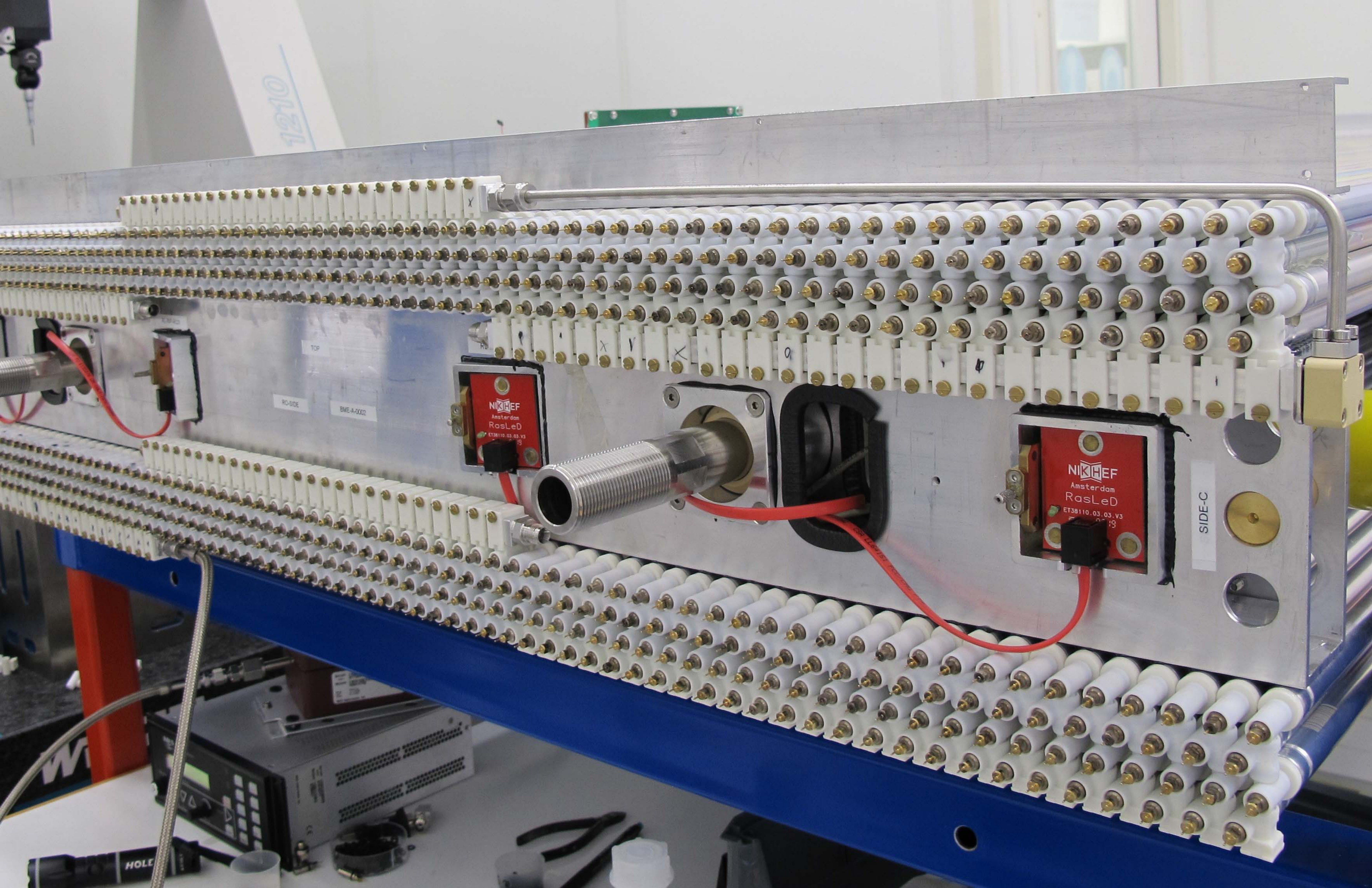}
\caption{Photograph of an assembled BME chamber with gas distribution system to the tubes mounted.}
\label{fig:BMEphoto}
\end{figure}

\vspace{-3mm}
Each layer of a BME sMDT chamber consists of 78 drift tubes of 2150~mm length.
Two BME chambers have been constructed in February and March 2014 and installed in April 2014. They are now taking data in the ATLAS experiment
at a center-of-mass energy of 13~TeV.
The sense wire positions in the BME chambers have been measured with a coordinate measuring machine (CMM) immediately after assembly.
The residual distributions with respect to the expected wire grid (see Fig.~\ref{fig:BMEA_wirepos}) show an unprecedentedly high wire positioning 
accuracy of better than $10~\mu$m in both transverse coordinates. 
The layout of a BME sMDT chamber with two RPC trigger chambers mounted on top and bottom is 
shown in Fig.~\ref{fig:BMEdesign}. Fig.~\ref{fig:BMEphoto} shows an assembled chamber with the gas distribution system installed.
Like the MDT chambers, the BME chambers carry in-plane optical monitoring systems for the measurement of torsion and gravitational 
sag of the chambers and to adjust the latter to the wire sagitta. 

\vspace{-5mm}
\section{BMG Chambers}

\begin{figure}
\centering
\includegraphics[width=0.7\linewidth]{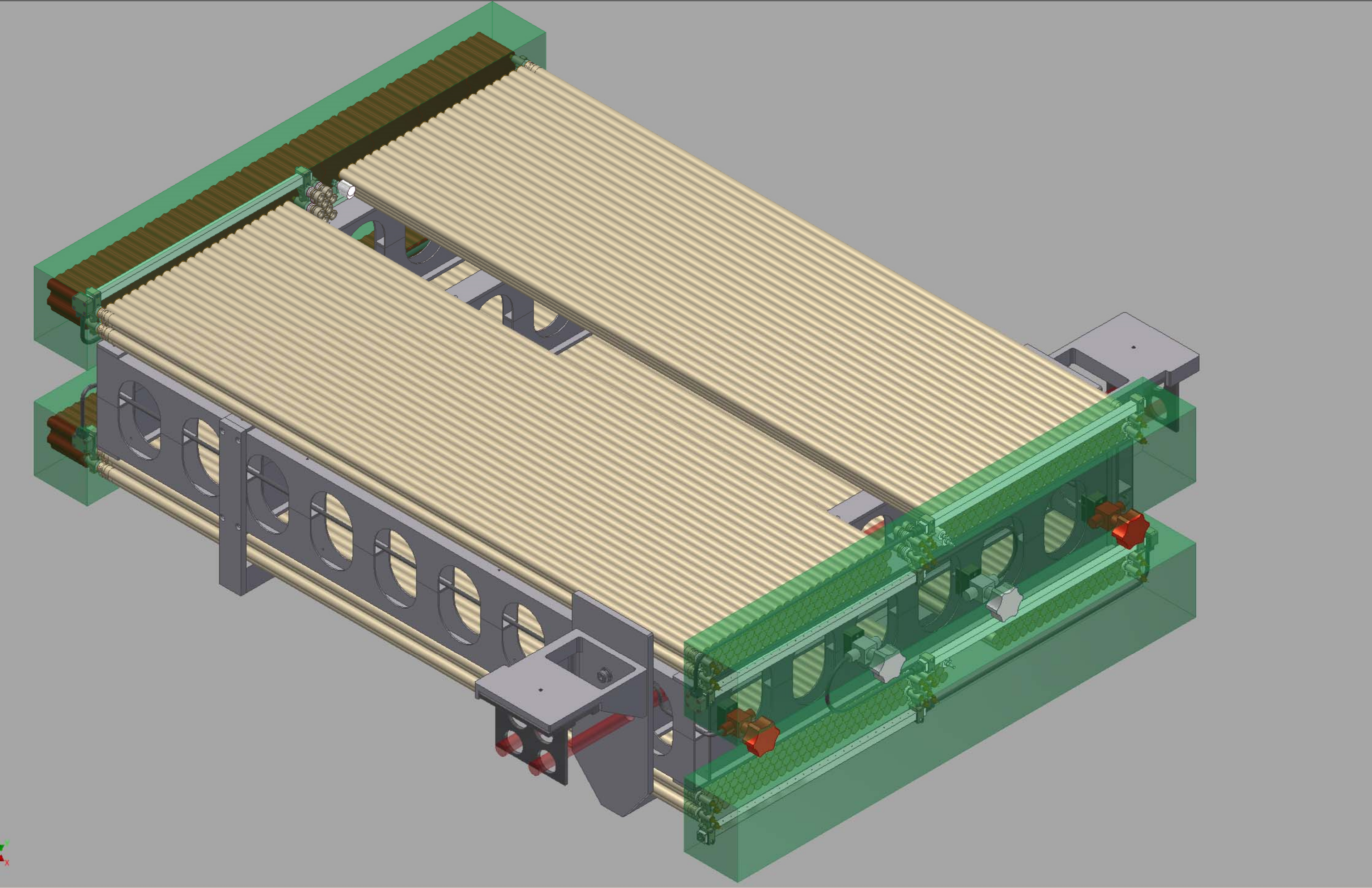}
\caption{Layout of a BMG chamber with extension platforms for alignment sensors.}
\label{fig:BMG_design}
\end{figure}

\begin{figure}
\centering
\vspace{-20mm}
\includegraphics[width=0.65\linewidth]{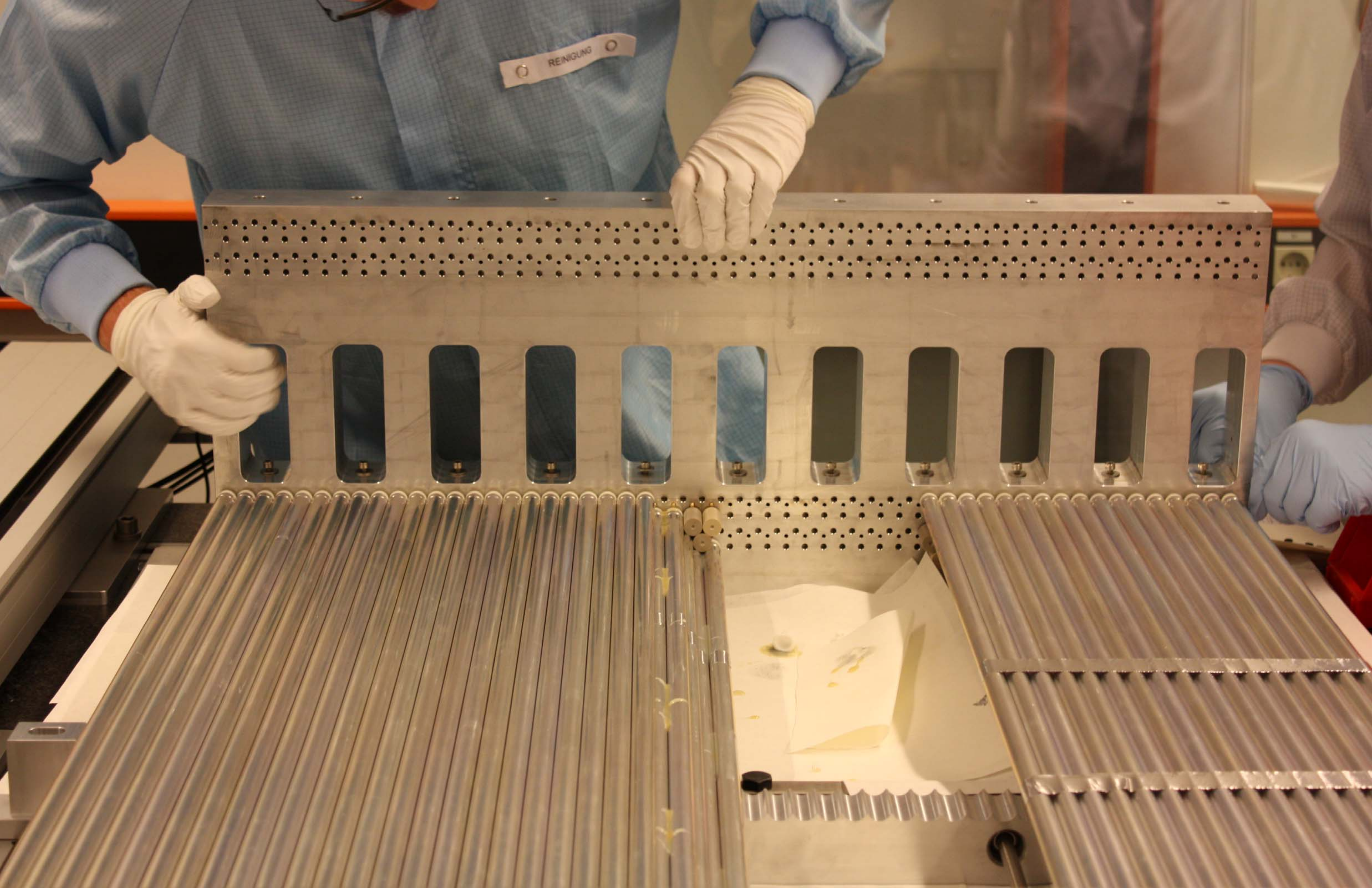}
\vspace{-1mm}
\caption{Assembly of a BMG chamber with the jigs with a hole grid for precise positioning of the endplugs of the drift tubes.}
\label{fig:BMGphoto}
\end{figure}

\vspace{-2mm}
BMG sMDT chambers consist of 54 drift tubes of 1129~mm length per layer.
The spacer frame is stiff enough to keep deformations below $20~\mu$m without the need for an optical
in-plane alignment system. The chambers have cutouts in the tube layers to let light rays of the
ATLAS optical alignment system pass (see Fig.~\ref{fig:BMG_design}). The first BMG chamber has already been 
assembled (see Fig.~\ref{fig:BMGphoto}).

\vspace{-3mm}
\section{Conclusions}

\vspace{-3mm}
Small-diameter Muon Drift-Tube (sMDT) chambers are very well suited for upgrades of the ATLAS muon spectrometer
at high LHC luminosities. Several upgrade projects with sMDT chambers are in progress in ATLAS. 
The new chambers show a factor of two higher mechanical accuracy, in spite of the simplified construction
procedures compared to the ATLAS MDT chambers, and an order of magnitude higher rate capability.

\vspace{-4mm}

\end{document}